\documentclass[conference, a4paper]{IEEEtran}
\IEEEoverridecommandlockouts
\usepackage{cite}
\usepackage{amsmath,amssymb,amsfonts}
\usepackage{algorithmic}
\usepackage{comment}
\usepackage{comment,threeparttable,multirow,array}
\usepackage{enumitem}
\usepackage{graphicx}
\usepackage{hhline}
\usepackage{textcomp}
\usepackage{xcolor}

\usepackage{subfig}
\newcommand{\vect}[1]{\boldsymbol{#1}}
\def\BibTeX{{\rm B\kern-.05em{\sc i\kern-.025em b}\kern-.08em
    T\kern-.1667em\lower.7ex\hbox{E}\kern-.125emX}}

\begin{document}
\title{On the Security of Pixel-Based Image Encryption for Privacy-Preserving Deep Neural Networks}
\author{\IEEEauthorblockN{Warit Sirichotedumrong, Yuma Kinoshita and Hitoshi Kiya} 
\IEEEauthorblockA{Tokyo Metropolitan University, Asahigaoka, Hino-shi, Tokyo, 191-0065, Japan}
\thanks{This work was partially supported by Grant-in-Aid for Scientific
Research(B), No.17H03267, from the Japan Society for the Promotion Science.} }

\maketitle
\begin{abstract}
This paper aims to evaluate the safety of a pixel-based image encryption method, which has been proposed to apply images with no visual information to deep neural networks (DNN), in terms of robustness against ciphertext-only attacks (COA). 
In addition, we propose a novel DNN-based COA that aims to reconstruct the visual information of encrypted images. 
The effectiveness of the proposed attack is evaluated under two encryption key conditions: same encryption key, and different encryption keys. 
The results show that the proposed attack can recover the visual information of the encrypted images if images are encrypted under same encryption key. 
Otherwise, the pixel-based image encryption method has robustness against COA.
\end{abstract}
\section{Introduction}

The spread of deep neural networks (DNNs) has greatly contributed
to solving complex tasks for many applications\cite{Donahue2014,Krizhevsky2012},
such as for computer vision, biomedical systems, and information technology.
Deep learning utilizes a large amount of data to extract 
representations of relevant features, so the performance is significantly improved\cite{Tishby2015,michael2018on}.
However, there are security issues when using deep learning in cloud environments to train and test data, 
such as data privacy, data leakage, and unauthorized data
access. Therefore, privacy-preserving DNNs have become an urgent challenge.

Various methods have been proposed for privacy-preserving computation. The methods are classified into two types: perceptual
encryption-based\cite{apsipa_svm, tanaka,Ito_2009,Tang_2014,kurihara2015encryption,KURIHARA2015,CHUMAN2017ICASSP,CHUMAN2017IEICE,ChumanIEEETrans,sirichotedumrong_kiya_2019,crypto_compress,2016_Gaata} and homomorphic encryption (HE)-based
\cite{Araki2016,Araki2017,Lu2016UsingFH,Aono2015,Shokri2015,Phong2018,Phong2018activation,cryptonets,Wang_ISCAS2018}. 
HE-based methods are the most secure options for
privacy preserving computation,
but they are applied to only limited DNNs\cite{Shokri2015,Phong2018,Phong2018activation,cryptonets,Wang_ISCAS2018}.
Therefore, the HE-based type does not support state-of-the-art DNNs yet.
Moreover, data augmentation has to be done before encryption.
In contrast, perceptual encryption-based methods
have been seeking a trade-off in security to enable other requirements, such
as a low processing demand, bitstream compliance, and signal processing
in the encrypted domain\cite{apsipa_svm, tanaka,Ito_2009,Tang_2014,kurihara2015encryption,KURIHARA2015,CHUMAN2017ICASSP,CHUMAN2017IEICE,sirichotedumrong_kiya_2019,ChumanIEEETrans,crypto_compress,2016_Gaata}.
A few methods were applied to machine learning algorithms in previous works\cite{apsipa_svm, tanaka}.
The first encryption method\cite{kurihara2015encryption,KURIHARA2015,CHUMAN2017ICASSP,CHUMAN2017IEICE,sirichotedumrong_kiya_2019,ChumanIEEETrans}
to be proposed for encryption-then-compression (EtC)
systems, was demonstrated to be applicable to traditional machine
learning algorithms, such as support vector machine (SVM)\cite{apsipa_svm}.
However, the block-based encryption method has never been applied to DNNs.

Although a block-based encryption method\cite{tanaka} was applied to image
classification with DNNs, in which an adaption network is added prior
to DNNs to avoid the influence of image encryption, the classification performance is inadequate.
A pixel-based image image encryption method\cite{WaritICIP2019} was proposed not only to improve the classification performance of the privacy-preserving DNNs but also to consider data augmentation in the encrypted domain.
However, the security level of these encryption methods is only evaluated in terms of the key space analysis for brute-force attack.

In this paper, we aim to evaluate the safety of the pixel-based image encryption in terms of robustness against ciphertext-only attacks (COA), including a new attack called DNN-based ciphertext-only attack.
Moreover, the effectiveness of the proposed attack is evaluated under two encryption key conditions: same encryption key, and different encryption keys. 

\section{Security Evaluation of Pixel-Based Image Encryption}

\subsection{Privacy-Preserving Deep Neural Networks}
\begin{figure}[t]
\centering
\includegraphics[width =8.5cm]{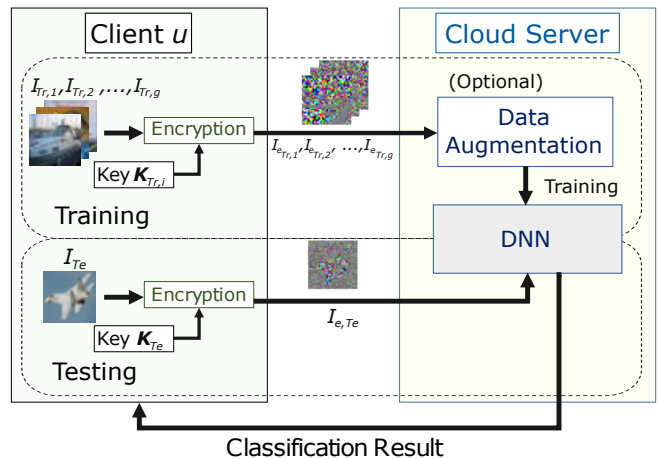}
\caption{Scenario}
\label{fig:scenario}
\end{figure}
Figure\,\ref{fig:scenario} illustrates the scenario for privacy-preserving DNNs used in this paper. In the training process, a client $u$ encrypts each training image, $I_{Tr,i}$, $i=1, 2,\ldots,g$, by using a secret key set for training data, $\vect{K}_{Tr,i}$, and sends the encrypted images ($I_{e_{Tr,i}}$) to a cloud server.

In the testing process, the client $u$ encrypts a testing image ($I_{Te}$) by using a secret key set for testing data, $\vect{K}_{Te}$, and sends the encrypted image $I_{e,Te}$ to the server.
The server solves a classification problem with an image classification model trained in advance, and then returns the classification results to the client.

In this paper, we assume that there are two encryption key conditions for generating encrypted images as follows.
\begin{itemize}[nosep]
  \item \textbf{\textit{Same encryption key:}} All training and testing images are encrypted by using only one secret key, i.e. $\vect{K}_{Tr,1}=\vect{K}_{Tr,2}=\ldots=\vect{K}_{Tr,g}=\vect{K}_{Te}=\vect{K}$.
  \item \textbf{\textit{Different encryption keys:}} The different secret keys are randomly assigned to training and testing images, i.e. $\vect{K}_{Tr,1}\neq\vect{K}_{Tr,2}\neq\ldots\neq\vect{K}_{Tr,g}\neq\vect{K}_{Te}$.
\end{itemize}

\begin{figure}[t]
\centering
\includegraphics[width =8.5cm]{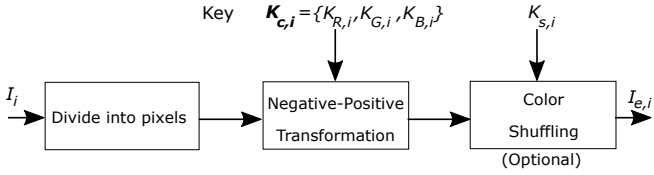}
\caption{Pixel-based image encryption}
\label{fig:encryption}
\end{figure}
\begin{figure}[!t]
\captionsetup[subfigure]{justification=centering}
\centering
\subfloat[Original image]{\includegraphics[clip, height=2.7cm]{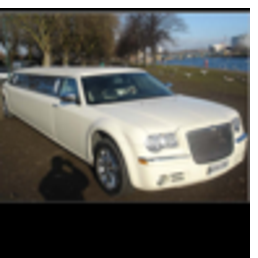}
\label{fig:label-A}}
\hfil
\subfloat[Negative-positive transformation]{\includegraphics[clip, height=2.7cm]{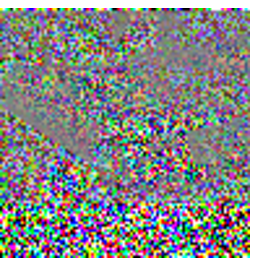}
\label{fig:label-B}}
\hfil
\subfloat[Negative-positive transformation and color shuffling]{\includegraphics[clip, height=2.7cm]{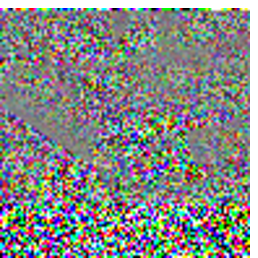}
\label{fig:label-C}}
\caption{Examples of images encrypted by proposed method}
\label{fig:proposed_images}
\end{figure}

\subsection{Pixel-Based Image Encryption}
Figure\,\ref{fig:encryption} illustrates the encryption steps of the pixel-based image encryption\cite{WaritICIP2019}.
To generate an encrypted image ($I_{e,i}$) from a color image, $I_{i}$, the following steps are carried out, as shown in Fig.\,\ref{fig:encryption}. Note that the color shuffling (Step 3) is an optional encryption step to enhance security.
\begin{itemize}
  \item [1)] Divide $I_{i}$ with $X \times Y$ pixels into pixels.
  \item [2)] Individually apply negative-positive transformation to each pixel of each color channel, $I_{R,i}$, $I_{G,i}$, and $I_{B,i}$, by using a random binary
   integer generated by secret keys $\vect{K_{c,i}}=\{K_{R,i}, K_{G,i}, K_{B,i}\}$. In this step, a transformed pixel value of the $j$-th pixel, $p'$, is calculated using 

	\begin{equation}
	\label{eq:p_nega-pos}
	p'=
	\left\{
	\begin{array}{ll}
	p & (r(j)=0) \\
	p \oplus (2^L-1) & (r(j)=1)
	\end{array} ,
	\right.
	\end{equation}
	where $r(j)$ is a random binary integer generated by $\vect{K_{c,i}}$. $p$ is
	the pixel value of the original image with $L$ bit per pixel. The
value of the occurrence probability $P(r(j))=0.5$ is used to invert bits
randomly\cite{ChumanIEEETrans}.

  \item [3)] (Optional) Shuffle three color components of each pixel by using an integer randomly selected from six integers 
  generated by a key $K_{s,i}$ as shown in Table\,\ref{tbl:color_shuf}.
\end{itemize}

Images encrypted by using the pixel-based method are illustrated in Fig.\,\ref{fig:proposed_images}(b) and \ref{fig:proposed_images}(c), where Fig.\,\ref{fig:proposed_images}(a) is the original one.
\begin{table}
\centering
\caption{Permutation of color components for random
integer. For example, if random integer is
equal to 2, red component is replaced by green
one, and green component is replaced by red one while blue component is not
replaced.}
\label{tbl:color_shuf}%%%Table caption goes heree
\begin{tabular}{|>{\centering\arraybackslash}m{1in}||>{\centering\arraybackslash}m{1cm}|>{\centering\arraybackslash}m{1cm}|>{\centering\arraybackslash}m{1cm}|}%%%The
% number of columns has to be defined here
\hline
\multirow{2}{*}{Random Integer} & \multicolumn{3}{c|}{Three Color Channels
}\\
\hhline{~---}
& R & G & B\\
 %%%% Table body
\hline
0 & R & G & B\\
\hline
1 & R & B & G\\
\hline
2 & G & R & B\\
\hline
3 & G & B & R\\
\hline
4 & B & R & G\\
\hline
5 & B & G & R\\
\hline
\end{tabular}
\end{table}%%%End of the table
\section{Robustness against Ciphertext-only Attacks}
Security mostly refers to protection from adversarial forces. Various attacking strategies, such as the known-plaintext attack (KPA) and chosen-plaintext attack (CPA), should be considered\cite{ChumanIEEETrans,CHUMAN2017ICASSP,CHUMAN2017IEICE}. 
In this paper, we consider brute-force attacks and propose a novel DNN-based ciphertext-only attack as ciphertext-only attacks (COA).
\subsection{Brute-force Attack}
If a color image $I_{RGB}$ with $X \times Y$ pixels is divided into pixels, the number of pixels $n$ is given by
\begin{equation}
\label{eq:blocknum}
n = X \times Y.
\end{equation}

The key spaces of negative-positive transformation ($N_{np}$) and color component shuffling ($N_{col}$) are represented by
\begin{equation}
{N_{np}(n)} = 2^{3n}, {N_{col}(n)} = \bigl( { }_3 P { }_3 \bigr)^{n} = 6^{n}.
\end{equation}

Consequently, the key space of images encrypted by using the proposed encryption scheme, $N(n)$, is represented by the following.
\begin{equation}
\begin{array}{ll}
N(n) & = {N_{np}(n)} \cdot {N_{col}(n)}
\\
& =2^{3n} \cdot 6^{n}
\end{array}
\end{equation}

\subsection{DNN-based Ciphertext-only Attack}
\begin{figure}[t]
\centering
\includegraphics[width =8.5cm]{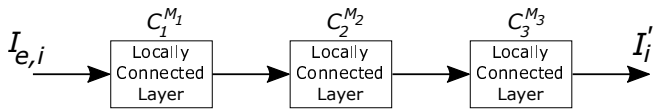}
\caption{Proposed DNN-based plaintext-only attack}
\label{fig:attack}
\end{figure}

We propose a novel DNN-based COA that aims to reconstruct the visual information of encrypted images.
 Since the encryption method is a pixel-based one, the proposed DNN for COA consists of three 1$\times$1-locally connected layers, which work similarly to 1$\times$1-convolution layer, except that weights are unshared. 
 Figure\,\ref{fig:attack} illustrates the proposed attack where $C_{j}^{M_j}$ is the $j$-th locally connected layer of the network with a kernel size and stride of (1,1),
  $M_j$ is the number of feature maps of the $j$-th locally connected layer, $j\in\{1,2,3\}$, and $I_i'$ denotes a reconstructed image.
The representations of each encrypted pixel are extracted in the first two layers, and then the reconstructed pixels are obtained by the last layer.
\begin{figure}[!t]
\captionsetup[subfigure]{justification=centering}
\centering
\subfloat[Same encryption keys]{\includegraphics[clip, height=3cm]{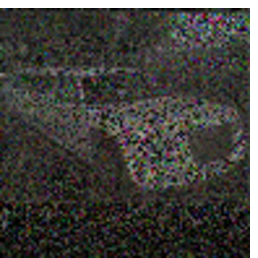}
\label{fig:label-A}}
\hfil
\subfloat[Different encryption key]{\includegraphics[clip, height=3cm]{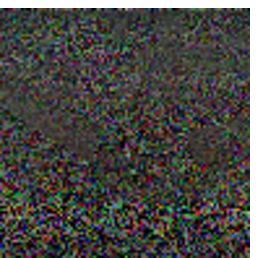}
\label{fig:label-B}}
\caption{Examples of reconstructed images from the images encrypted by the negative-positive transformation and color shuffling.}
\label{fig:recon}
\end{figure}
\begin{table}
\small
\centering
\caption{Average SSIM of the reconstructed images compared with the original ones.}

\label{tbl:result}%%%Table caption goes heree
\begin{tabular}{|c|c|c|}%%%The
% number of columns has to be defined here
\hline
Key Conditions & Encryption & SSIM\\
\hline
\multirow{2}{*}{Same encryption key}&Step 2&\textbf{0.1732}\\
\hhline{~--}
&Step 2 and 3& 0.1715\\
\hline
\multirow{2}{*}{Different encryption keys}&Step 2&0.0424\\
\hhline{~--}
&Step 2 and 3& 0.0425\\
\hline
\end{tabular}

\end{table}%%%End of the table

\section{Experiments}

\subsection{Experimental Set-up}

We employed STL-10 dataset, which contains $96 \times 96$ pixel color images and consists of 5,000 training images and 8,000 testing images\cite{stl10}.

In the experiment, the numbers of feature maps, $M_1$, $M_2$, and $M_3$, were set to 8, 32, and 3, respectively.

The network was trained by using stochastic gradient descent (SGD) with momentum for 70 epochs, and used mean squared error (MSE), which compares the differences between the reconstructed images and the original ones, as a loss function. 
The learning rate was initially set to 0.1 and decreased by a factor of 10 at 40 and 60 epochs. 
We used a weight decay of 0.0005, a momentum of 0.9, and a batch size of 128.

\subsection{Results and Discussions}

Examples of reconstructed images under the use of same and different encryption keys are shown in Fig.\,\ref{fig:recon}, where Fig.\,\ref{fig:proposed_images}(a) is the original one. 
The visual information of the reconstructed images was recovered by the proposed scheme if the images are encrypted under same encryption key, as shown in Fig.\,\ref{fig:recon}(a). This is because each image was encrypted with only one pattern, so the proposed attack can recognize the pattern and recover the visual information by comparing the difference between reconstructed images and original images. 
In comparison, the pixel-based encryption method has robustness against COA if the training images are encrypted by using different encryption keys. Therefore, the visual information cannot be recovered, as shown in Fig.\,\ref{fig:recon}(b).

Table\,\ref{tbl:result} shows that the structural similarity (SSIM) values of the encrypted images under the use of same encryption key were much higher than under the use of different encryption keys.

\section{Conclusion}

This paper aimed to evaluate the safety of the pixel-based image encryption method in terms of robustness against COA. 
In addition, we proposed a novel DNN-based COA that aims to reconstruct the visual information of encrypted images. 
The effectiveness of the proposed attack was evaluated under two encryption key conditions: same encryption key, and different encryption keys.
The experimental results showed that the proposed attack can recover the visual information of the encrypted images if images are encrypted under same encryption key.
In contrast, it was proved that the pixel-based image encryption method has robustness against COA if images are encrypted under different encryption keys.
\bibliographystyle{IEEEtran}

\end{document}